\documentclass[%
 reprint,
superscriptaddress,
 amsmath,amssymb,
 aps,
prb,
]{revtex4-2}

\usepackage{graphicx}
\usepackage{dcolumn}
\usepackage{bm}
\usepackage{hyperref}
\usepackage{soul}

\usepackage{graphicx, multirow, xcolor, bm, ulem}
\usepackage{amsfonts,amssymb,amsmath}
\usepackage{graphicx,dcolumn,bm,color,braket,slashed}
\usepackage{times, comment, mathtools, extarrows,textcomp,ulem}

\begin{document}

\title{Field-Induced Selective Spin Gap Closure and Quantum Criticality in BaNd$_2$ZnS$_5$}

\author{Sangyun Lee}
\email[Contact author: ]{sangyun.lee@ufl.edu}
\affiliation{National High Magnetic Field Laboratory, Los Alamos National Laboratory, Los Alamos, New Mexico 87545, USA.}
\affiliation{Department of Physics and National High Magnetic Field Laboratory High B/T Facility, University of Florida, Gainesville, Florida 32611-8440, USA.}

\author{A. J. Woods}
\affiliation{MPA-Q, Los Alamos National Laboratory, Los Alamos, NM 87545, USA.}
\affiliation{National High Magnetic Field Laboratory, Tallahassee, Florida 32310, USA.}

\author{B. Billingsley}
\affiliation{Department of Physics, University of Arizona, Tucson, Arizona, 85721, USA.}

\author{Shengzhi Zhang}
\affiliation{National High Magnetic Field Laboratory, Los Alamos National Laboratory, Los Alamos, New Mexico 87545, USA.}

\author{R. Movshovich}
\affiliation{MPA-Q, Los Alamos National Laboratory, Los Alamos, NM 87545, USA.}

\author{S. M. Thomas}
\affiliation{MPA-Q, Los Alamos National Laboratory, Los Alamos, NM 87545, USA.}


\author{C. A. Mizzi}
\affiliation{National High Magnetic Field Laboratory, Los Alamos National Laboratory, Los Alamos, New Mexico 87545, USA.}

\author{B. Maiorov}
\affiliation{National High Magnetic Field Laboratory, Los Alamos National Laboratory, Los Alamos, New Mexico 87545, USA.}

\author{Shuyi Li}
\affiliation{Department of Physics, University of Florida, Gainesville, Florida 32611-8440, USA.}

\author{Chunjing Jia}
\affiliation{Department of Physics, University of Florida, Gainesville, Florida 32611-8440, USA.}

\author{Tai Kong}
\affiliation{Department of Physics, University of Arizona, Tucson, Arizona, 85721, USA.}
\affiliation{Department of Chemistry and Biochemistry, University of Arizona, Tucson, Arizona, 85721, USA.}

\author{Eun Sang Choi}
\affiliation{National High Magnetic Field Laboratory, Tallahassee, Florida 32310, USA.}

\author{Vivien S. Zapf}
\affiliation{National High Magnetic Field Laboratory, Los Alamos National Laboratory, Los Alamos, New Mexico 87545, USA.}

\author{Minseong Lee}
\email[Contact author: ]{ml10k@lanl.gov}
\affiliation{National High Magnetic Field Laboratory, Los Alamos National Laboratory, Los Alamos, New Mexico 87545, USA.}

\begin{abstract}
We report thermodynamic evidence for field-induced mode-selective quantum criticality in the layered rare-earth magnet BaNd$_2$ZnS$_5$ (BNZS). Below the Néel temperature $T_N = 2.9$ K, spin–orbit–entangled Nd$^{3+}$ moments form two symmetry-inequivalent low-energy spin-excitation modes arising from Kramers doublet physics under a magnetic field, with distinct gaps $\Delta_L$ and $\Delta_H$. For magnetic fields applied along the [110] direction, the lower-energy gap $\Delta_L$ softens continuously and collapses at a critical field $H_c \approx 2$ T, while the higher-energy gap $\Delta_H$ remains gapped, leaving the system in an intermediate partially critical phase. Despite the partial nature of the criticality, thermodynamic measurements reveal a continuous quantum phase transition. The ac susceptibility shows universal scaling behavior, with $\chi_{ac}(T, H)$ collapsing onto a single scaling function and following $\chi_{ac} \sim T^{-0.2}$ at criticality. A finite residual Sommerfeld coefficient $\gamma_0$ further indicates the emergence of gapless excitations confined to a single symmetry sector near the quantum critical point. In contrast to conventional quantum criticality based on global softening of low-energy excitations, BNZS exhibits a selective breakdown of Kramers-doublet excitations due to its strong anisotropic interactions. Our results establish BNZS as a spin–orbit–coupled rare-earth magnet where quantum criticality is not global but mode-selective, with anisotropic interactions enabling independent criticality in distinct excitation sectors.
\end{abstract}

\maketitle

\clearpage


In quantum magnets, the nature of the ground state is intimately connected to the spectrum of low-energy excitations. Accordingly, phase transitions accompanied by changes in the spin structure of the ground state often lead to a global reconstruction of the low-energy excitation spectrum. In systems with predominantly isotropic interactions, this reconstruction is typically collective, involving a broad reorganization of spin excitations across the system~\cite{Sachdev2011,Vojta2003,Balents2010,Coldea2010}. In contrast, quantum magnets with strong spin–orbit entanglement provide a fundamentally different platform~\cite{Gohlke2018,Yuji2025,Zhou2017,kumar2024,samanta2024}. The interplay between spin–orbit coupling and crystal electric field generates highly anisotropic exchange interactions~\cite{kugel1982jahn,jackeli2009mott,witczak2014correlated,rau2016spin}, which can stabilize exotic quantum ground states, including quantum spin liquid–like regimes~\cite{banerjee2017neutron,kasahara2018majorana,baek2017evidence}. Moreover, they can effectively project spin interactions onto specific directions in spin space, allowing only selected spin components or modes to become critical across a quantum phase transition induced by magnetic fields. This opens the possibility of mode-selective criticality, where only a subset of low-energy excitations becomes critical, instead of a global reconstruction of the low-energy excitation spectrum ~\cite{Sondhi1997,Gegenwart2008,BECKhatua2025,BECMukharjee2019,BECRadu2005,BECZhao2022}.
Such mode-selective quantum criticality enables intriguing quantum states in which bosonic excitations associated with spin ordering coexist with exotic fermionic—or even non–quasiparticle—excitations emerging from a quantum critical sector ~\cite{Si2001,Senthil2003,Gegenwart2008}. This, in turn, raises the possibility of interacting degrees of freedom that obey distinct quantum statistics within the same system. Despite this intriguing potential, the underlying physical properties of such states remain largely unexplored, including the nature of the phase transition, the associated critical exponents, and the universality of the critical behavior. 

In this Letter, we report thermodynamic evidence for a mode-selective quantum critical point in the layered rare-earth chalcogenide BaNd$_2$ZnS$_5$ (BNZS) by probing the field-induced evolution of spin excitations using magnetization, ac susceptibility, heat capacity, magnetocaloric effect, and ultrasound measurements. Our measurements show that only one of the two order parameters collapses, as revealed by the phase diagram and supported by a theoretical model. The scaling of the ac susceptibility further demonstrates that the quantum phase transition is continuous.

BNZS possesses Shastry–Sutherland-type layers, although its magnetic properties are distinct from those of canonical Shastry–Sutherland systems \cite{Billingsley2022}. It orders antiferromagnetically below $T_{\mathrm{N}} = 2.9$ K. Neutron diffraction measurements show that BNZS hosts a magnetic structure characterized by two propagation vectors $q_{1}$ and $q_{2}$ that coexist at zero field \cite{Marshall2022}. The presence of two orthogonal propagation vectors suggests at strong anisotropic exchange and the existence of multiple magnetic modes. Neutron diffraction further reveals that only the $q_{2}$ mode is suppressed under a magnetic field applied along the crystallographic [110] direction near 2 T, while $q_{1}$ remains robust. Consistently, magnetization measurements exhibit strong anisotropy for fields applied along different crystallographic directions, indicating a pronounced $g$-factor anisotropy arising from strong spin–orbit coupling \cite{Billingsley2022}.

\begin{figure}
\centering
\includegraphics[width=1\columnwidth]{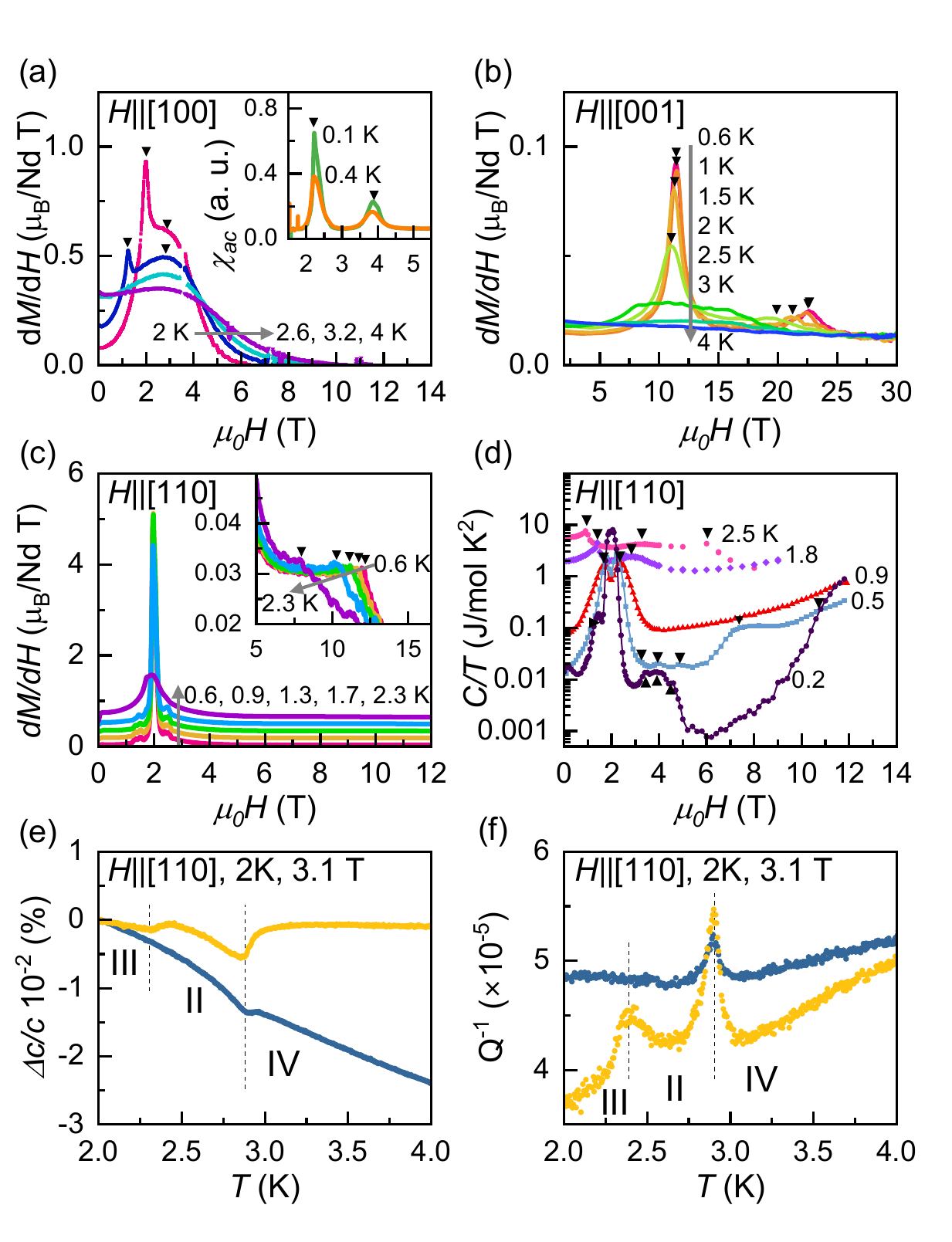}
\caption{\label{fig:fig1L} First derivative of the dc magnetization ($dM/dH$) as a function of magnetic field for (a) $H||[100]$, (b) $H||[001]$, and (c) $H||[110]$. Phase boundaries were determined from the peaks in $dM/dH$. The inset of (a) displays the ac magnetic susceptibility ($\chi_{\mathrm{ac}}$) versus magnetic field at 0.1 and 0.4 K. The inset of (c) shows an expanded view of $dM/dH$ near 12~T. Panel (d) presents the specific heat divided by temperature ($C/T$) as a function of magnetic field for $H \parallel [110]$ at various temperatures. (e,f) Temperature-dependent stiffness and attenuation at 3.1~T for $H\parallel[110]$ measured by resonant ultrasound spectroscopy (RUS). Different resonances show distinct responses across the phase transitions, indicating a symmetry-selective response.}
\end{figure}

Fig.~\ref{fig:fig1L} (a--c) show the field derivative of the dc magnetization, $dM/dH$, measured with magnetic fields applied along the [100], [001], and [110] crystallographic directions.
The ac susceptibility was measured using an excitation field $H_{ac}=3.4$~Oe at a frequency of 271~Hz, with $H_{ac}$ applied parallel to the dc field $H$. The ac susceptibility method is described in detail in \cite{dun2014chemical,lee2016magnetic}.
For $H\parallel[100]$ [Fig.~\ref{fig:fig1L}~(a)], two distinct peaks in $dM/dH$ are observed at $T=2.6$~K, which sharpen upon cooling.
As shown in the inset, these anomalies evolve into sharp transitions at $H = 2.2$~T and 3.9~T for $T=0.4$~K.
No additional transitions are detected up to 18~T, consistent with magnetization saturation and stabilization of two field-induced phases for this field orientation in addition to the zero field phase.

For $H\parallel[001]$ [Fig.~\ref{fig:fig1L}(b)], a sequence of three phases is also observed, marked by pronounced peaks in $dM/dH$ at 11.6~T and 22.6~T.
Pulsed-field measurements up to 60~T reveal no additional transitions.
Above 22.6~T, the magnetization increases linearly with field, consistent with a sizable Van Vleck contribution of approximately $0.01~\mu_B$/T per Nd arising from spin--orbit coupling and crystal-field effects, comparable to that reported in other Nd$^{3+}$-based compounds  \cite{xu2019anisotropic}.
The saturated moment for $H\parallel[001]$ ($M_s = 0.3~\mu_B$/Nd) is significantly smaller than that for $H\parallel[100]$ ($1.8~\mu_B$/Nd), rendering the linear Van Vleck contribution particularly prominent at high fields (see Fig. S11 and S14).

In contrast, measurements for $H\parallel[110]$ [Fig.~\ref{fig:fig1L}(c,d)] reveal a far more intricate sequence of field-induced transitions.
Both $dM/dH$ and the specific heat $C/T$ exhibit clear anomalies near 2~T and 12~T.
The high-field anomaly, highlighted in the inset around 12~T, corresponds to the gradual suppression of the N\'eel order, which completely vanishes at 12.7~T.
Consistent with previous neutron diffraction measurements~\cite{Marshall2022}, these results indicate that the $q_2$ magnetic component collapses near 2~T, whereas the $q_1$ order remains robust up to 12.7~T.

At lower temperatures, the $H\parallel[110]$ data reveal an increasingly complex sequence of field-induced phases, reflecting successive symmetry-selective rearrangements of the magnetic state. 
Multiple anomalies develop and evolve with temperature, indicating a hierarchy of low-energy instabilities that sharpen upon cooling [Fig.~\ref{fig:fig1L}(d)].
Taken together, these observations establish that the field-driven evolution is not a simple suppression of magnetic order, but rather proceeds through a cascade of intermediate states before the eventual collapse of long-range order.

Temperature-dependent measurements of the elastic response further support this interpretation~\cite{Mizzi2024RUSHighField}. 
Figures~\ref{fig:fig1L}(e,f) show the stiffness and attenuation at 3.1~T for $H\parallel[110]$, measured using resonant ultrasound spectroscopy (RUS). 
The Phase~II–IV transition is observed in both resonances, whereas the Phase~III–II transition appears only in one mode, revealing a symmetry-selective coupling of the lattice to the underlying magnetic degrees of freedom. 
This selective response provides direct evidence that distinct symmetry channels are involved across the phase boundaries, reinforcing the picture of mode-selective critical behavior.

\begin{figure}
\centering
\includegraphics[width=1\columnwidth]{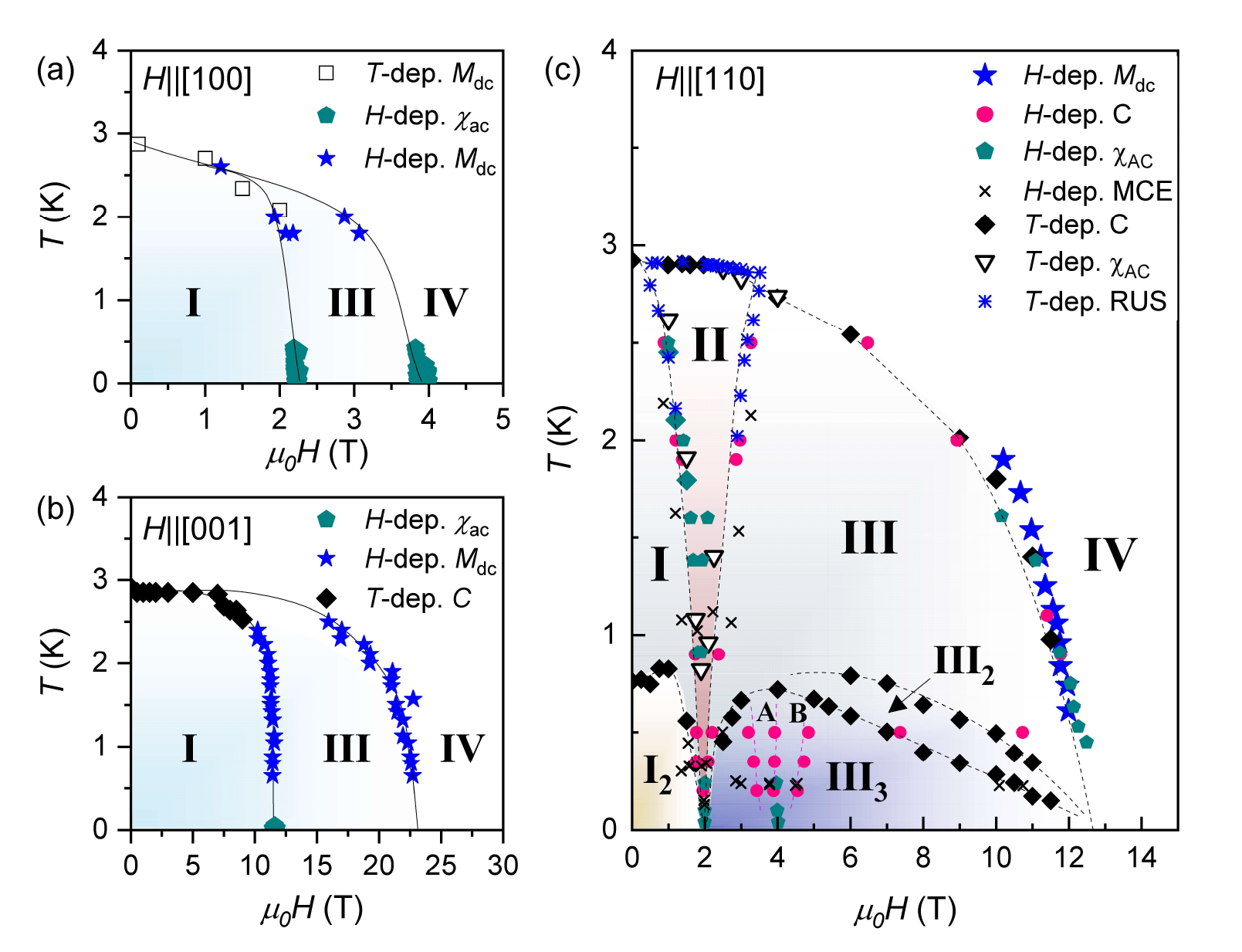}
\caption{\label{fig:phases} The magnetic phase diagram of BaNd$_2$ZnS$_5$ for $H||$[100] {\bf (a)}, $H||$[001] {\bf (b)}, and $H||$[110] {\bf (c)}. The phase diagram for $H||$[110] uniquely exhibits phase II, alongside other additional phases at low temperature, whereas the phase diagrams for other magnetic orientations only show phases I, III, and IV.}
\end{figure}

Fig.~\ref{fig:phases} summarizes the $H$--$T$ phase diagrams of BNZS for magnetic fields applied along the [100], [001], and [110] crystallographic directions (see Sections~S1.2--S1.4 of the Supplementary Information for details of the data used to determine the phase boundaries).
The long-range antiferromagnetic order below $T_N~=~2.9$~K defines Phase~I, which is characterized by the coexistence of two stripe-like magnetic modulations with propagation vectors $q_1$ and $q_2$, with moments lying predominantly in the $ab$ plane~\cite{Marshall2022}.

For $H\parallel[100]$ [Fig.~\ref{fig:phases}(a)], increasing magnetic field gradually suppresses $T_N$, and Phase~I disappears at a critical field $H_1 = 2.4$~T.
A field-induced Phase~III emerges for $H>H_1$ and persists up to $H_2 = 3.9$~T, above which the magnetization saturates.
Within Phase~III, the $q_1$ stripe order aligns along the field direction, while the $q_2$ modulation remains perpendicular, consistent with the underlying anisotropic exchange interactions.

A similar sequence is observed for $H\parallel[001]$ [Fig.~\ref{fig:phases}(b)], where Phase~I is successively replaced by Phase~III and then by Phase~IV, corresponding to magnetization saturation.
However, the characteristic fields are substantially larger, with $H_1 = 11.6$~T and $H_2= 22.6$~T.
The roughly fivefold enhancement of the critical fields compared to $H\parallel[100]$ scales with the reduced saturated moment for this orientation, highlighting the strong single-ion anisotropy of the Nd$^{3+}$ moments.

In contrast, the phase diagram for $H\parallel[110]$ [Fig.~\ref{fig:phases}(c)] exhibits a qualitatively different and richer structure.
Here, Phase~I evolves into an intermediate Phase~II, which is absent for the other field orientations. The region labeled Phase II corresponds to a partially critical regime in which long-range magnetic order associated with the $q_1$ sector coexists with enhanced low-energy excitations from the selectively softened $q_2$ branch.
Polarized neutron diffraction measurements show that the $q_1$ order remains robust up to $6$~T, while the $q_2$ component is suppressed below $2$~T, giving rise to a liquid-like magnetic response near the critical field $H_c \approx 2$~T~\cite{Marshall2022}.
Upon further increasing the field, Phase~II gives way to Phase~III, which persists up to $12.7$~T before the complete suppression of long-range magnetic order.

At temperatures below $1$~K, additional low-temperature complexity emerges. Phases I and III exhibit additional low-temperature features (denoted I$_i$ and III$_i$), which likely reflect changes in the excitation spectrum and may correspond to crossovers rather than symmetry-breaking transitions, and two additional regimes (A and B) appear within the Phase~III$_3$ dome near 4~T, as labeled in [Fig.~\ref{fig:phases}(c)].
The Phase~I subphases vanish at $2$~T, while the family of Phase~III states emerges at this field and ultimately collapses near 12.7~T.

Comparing all three field orientations, only $H\parallel[110]$ stabilizes an intermediate phase and multiple low-temperature splittings.
This distinction follows naturally from symmetry considerations: for $H\parallel[100]$ or $[001]$, the two modulation vectors $q_1$ and $q_2$ remain symmetry-equivalent under the applied field, leading primarily to shifts of phase boundaries.
In contrast, for $H\parallel[110]$, the external field breaks this equivalence, rendering the two magnetic components inequivalent and enabling the emergence of Phase~II.
These observations demonstrate that the phase complexity in BNZS arises from symmetry-selective magnetic coupling between the field and the two stripe modulations, which cannot be captured within a simple Ising-type description~\cite{Maksimov2019_PRX,Lee2025_npJQM}.

\begin{figure}
\centering
\includegraphics[width=1\columnwidth]{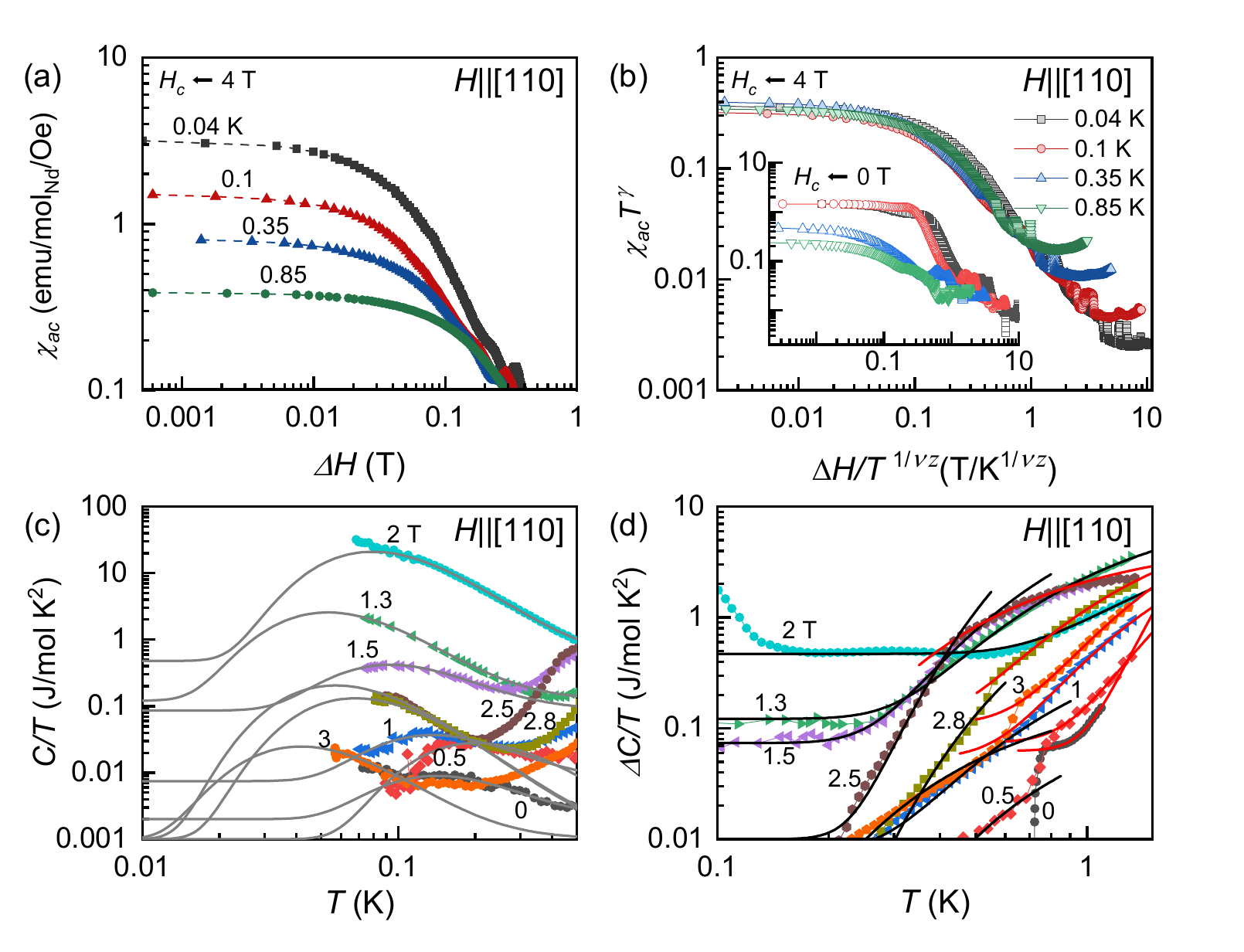}
\caption{\label{fig:qcp} ac magnetic susceptibility ($\chi_{ac}$) and specific heat divided by temperature ($C/T$) near 2~T for $H||$[110]. 
In {\bf (a)} $\chi_{ac}$ versus the rescaled magnetic field $\Delta H = H - H_c$, 
shown on a logarithmic scale for $H$ approaching $H_c$ from $4\,\mathrm{T}$. 
{\bf (b)} Scaling collapse of $\chi_{ac}$ obtained from (a) using critical exponents $\gamma = 0.67$ and $\nu z = 2.1$. 
Inset: Scaling analysis for $H \rightarrow H_c$ at $0\,\mathrm{T}$, where no data collapse is observed.
{\bf (c)} $C/T$ vs temperature under various magnetic field.
Rising of $C/T$ with decreasing at low temperature is shown two-level schottky anomaly fit (Gray lines). Details are in the main text.
{\bf (d)} $C/T$ vs temperature is subtracted by contribution of the schottky anomaly ($\Delta C/T$). Black and Red lines indicate fit to extract spin gaps from $\Delta C/T \sim (e^{-\Delta_{\text{spin}}/T})/T^n$ in temperature range below or above the kink (between I and I$_2$, or between III and III$_3$), respectively. 
}
\end{figure}

\begin{figure}
\centering
\includegraphics[width=1\columnwidth]{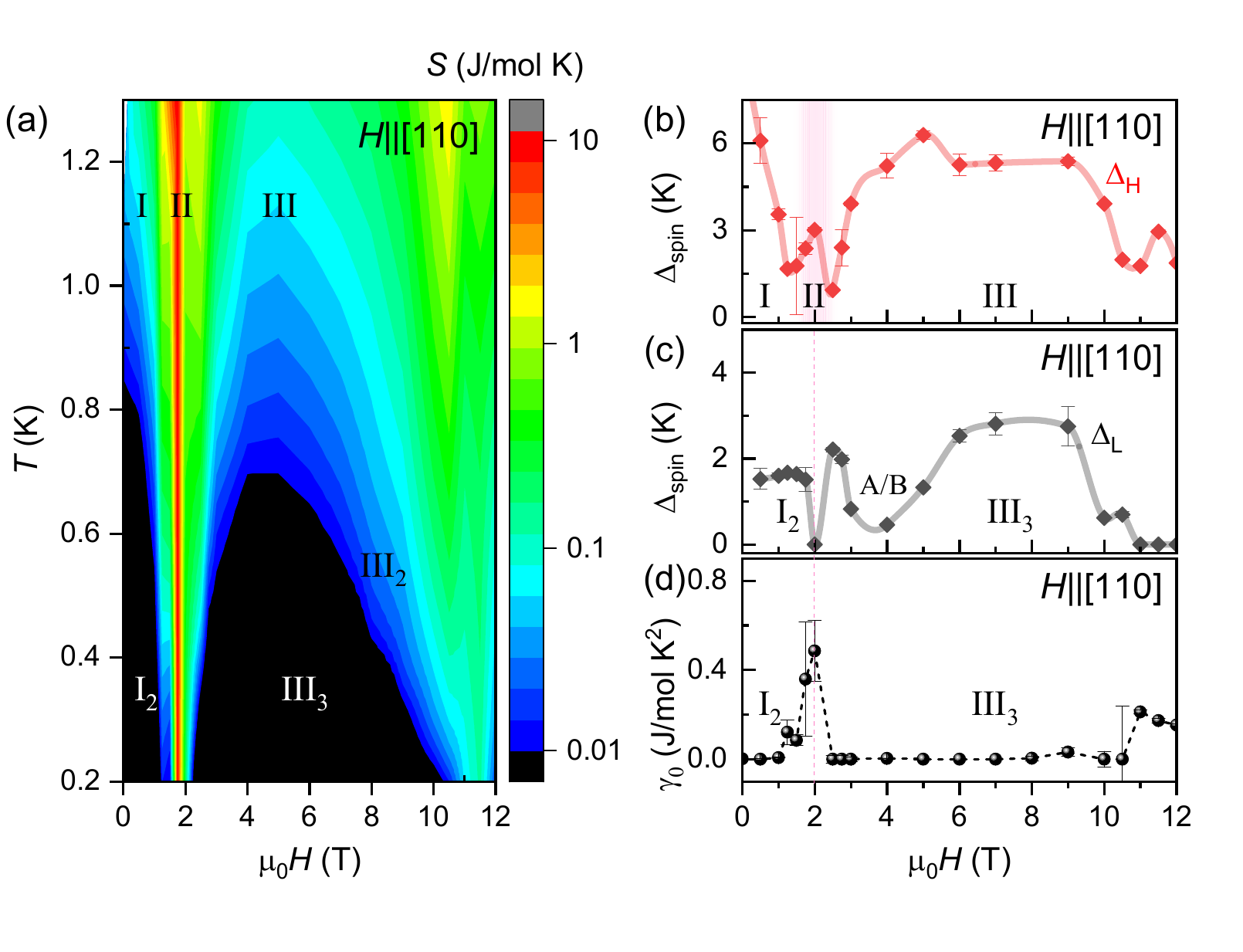}
\caption{\label{fig:phasecontour} {\bf (a)} The contour plot of entropy ($S$) estimated from integration of specific heat ($\Delta C/T$) in Fig. 3(d).
{\bf (b,c)} $\Delta_{\text{spin}}$ vs magnetic field for two different regime.
Red (Black) indicates extracted $\Delta_{\text{spin}}$ in temperature regime of phase I, II and III (phase I$_2$ and III$_3$) from the fits in Fig.3(d). $\Delta_L$ and $\Delta_H$ reflect spin gaps from I and I$_2$ ($H<2~T$) and III and III$_3$ ($H>2~T$), respectively.
Near 2~T, $\Delta_L$ becomes zero, while $\Delta_H$ is still preserved.
{\bf (d)} Residual Sommerfeld coefficient ($\gamma_0$) shows divergence near 2~T.
The pink shaded region and dashed lines indicate the critical field near 2~T.
Additional peak is pronounced near 12~T. 
}
\end{figure}

%

The most striking observation for $H||[110]$ is the emergence of quantum critical behavior near $H_c$.
Importantly, this criticality does not correspond to a complete loss of magnetic order but instead develops within Phase II, where the $q_1$ stripe order remains intact while only the $q_2$ sector becomes critical.
Peaks in both $C/T$ and $\chi_{ac}$ sharpen and move toward $T=0$, consistent with a continuous quantum phase transition. Thermodynamic consistency further supports this interpretation: while the Clausius--Clapeyron relation fails, the Ehrenfest relation yields quantitatively consistent results based on the observed continuous heat-capacity anomalies, supporting a second-order character of the transition within experimental resolution. [see Sections S2 in the Supplementary Information]~\cite{kumar1997properties,pippard1966thermodynamics}.

The magnetic susceptibility provides additional evidence for quantum criticality.
At $H_c$, $\chi_{ac}(T)$ exhibits a pronounced enhancement, increasing by a factor of approximately 3.6 between 0.04~K and 1~K. 
Over this temperature window, the data can be described by an effective power-law form $\chi_{ac}(T)=\chi_0+bT^{-\Gamma_T}$ with $\Gamma_T\approx0.2$.
In contrast, $\chi_{ac}(T)$ for $H||[100]$ remains non-divergent near $H_1$ or $H_2$, instead saturating or decreasing at low temperatures [see Fig.~S1].

We further examine the scaling behavior of the magnetic susceptibility near the critical field. 
Fig.~\ref{fig:qcp}(a) shows $\chi_{ac}$ plotted as a function of the rescaled field $\Delta H = H - H_c$ on a logarithmic scale for $H$ approaching $H_c$ from $4\,\mathrm{T}$. 
As the field approaches the critical value, $\chi_{ac}$ exhibits a clear power-law divergence, consistent with critical fluctuations associated with a quantum critical point~\cite{Gegenwart2008}.

To test the quantum critical scaling, the data are analyzed using the scaling form $\chi_{ac}(T,\Delta H) = T^{-\gamma} f(\Delta H / T^{1/(\nu z)})$, which consistently captures the broader critical regime~\cite{Sondhi1997}. 
As shown in Fig.~\ref{fig:qcp}(b), the data collapse onto a single universal curve with critical exponents $\gamma = 0.67$ and $\nu z = 2.1$, demonstrating robust quantum critical scaling in the vicinity of $H_c$. For comparison, the same scaling analysis performed for $H \rightarrow H_c$ at $0\,\mathrm{T}$ does not produce a scaling collapse (inset of Fig.~\ref{fig:qcp}(b)), indicating that the scaling behavior is specific to the field-tuned quantum critical regime.

The unusually small exponents $\Gamma_T \approx 0.2$ and $\gamma \approx 0.67$ indicate that criticality develops only within a restricted subset of the spin degrees of freedom, while the remaining excitation branch remains gapped and magnetically ordered.
This selective criticality gives rise to a sublinear divergence of the magnetic susceptibility, reflecting the absence of global softening of the low-energy excitation spectrum.
Such behavior cannot be captured within conventional descriptions based on magnetic Bose--Einstein condensation or Ising universality, which in their standard realization assumes a uniform collapse of all low-energy modes at the critical point~\cite{EXNikuni2000,EXHarrison2006}.

Figure~\ref{fig:qcp}(c) shows the temperature dependence of the specific heat.
As $T$ decreases, a pronounced hump appears, which is well accounted for by a two-level nuclear Schottky anomaly (solid gray line).
At $H_c \approx 2$~T, slight deviations at the lowest temperatures indicate contributions from a multi-level structure rather than a simple two-level model.
Fitting the data with $C/T = \gamma_0 + b_\mathrm{sch} C/T_\mathrm{Schottky}$, we find that the residual linear specific heat coefficient $\gamma_0$ increases markedly near $H_c$ [Fig.~\ref{fig:phasecontour}(d)].
Figure~\ref{fig:qcp}(d) presents the specific heat after subtraction of the Schottky contribution.
The red and black lines represent phenomenological fits using an activated exponential form, $\Delta C/T = \exp(-\Delta_\mathrm{spin}/T)/T^n$, in distinct temperature ranges corresponding to the high- and low-temperature regimes, i.e., Phase~I (III) and Phase~I$_2$ (III$_2$), respectively.
In the high-temperature regime, the extracted gap $\Delta_H$ remains finite across all fields, whereas in the low-temperature regime, $\Delta_L$ vanishes at $H_c$ [Fig.~\ref{fig:phasecontour}(b,c)].
Thus, the divergence of the magnetic susceptibility and the quantum critical point at $H_c$ occur exclusively for $H||[110]$, reflecting symmetry-selective softening of the $q_2$ excitation branch.

At higher temperatures (Phases~I and~III), both Kramers-doublet--derived excitation branches contribute to the heat capacity, giving rising to a large effective spin gap $\Delta_\mathrm{eff}\approx\Delta_H=9$~K in Phase~I ($\mu_{0}H = 0$), where the higher-energy branch remains thermally populated but gapped.
As the $q_2$ branch softens upon approaching the I/II boundary, $\Delta_H$ is reduced to approximately 1.5~K.
Within Phase~II, $\Delta_H$ exhibits a dome-like field dependence with a maximum value of about 3~K. Beyond Phase~II the $q_2$ excitations vanish, and the $q_1$ branch recovers a stable gap $\Delta_H\approx6$~K.

Upon cooling toward zero temperature (Phases~I$_2$ and~III$_2$), thermal fluctuations are suppressed and the intrinsic low-energy gaps $\Delta_L$ converge to common value of approximately 3~K across the field-tuned phases, reflecting a reorganization of the Kramers-doublet excitations. At the critical field $H_c$, however, $\Delta_L$ collapses and vanishes. 
The existence of two distinct low-energy scales is directly evident from the thermodynamic response and does not depend on the specific fitting form used to extract $\Delta_L$ and $\Delta_H$.
Further discussion based on a toy model is provided in Sections~S3--S5 of the Supplementary Information.


Taken together, the thermodynamic data summarized in Fig.~\ref{fig:phasecontour} provide a consistent picture of mode-selective quantum criticality. 
The entropy landscape $S(T,H)$ [Fig.~\ref{fig:phasecontour}(a)] exhibits a pronounced enhancement as $T \rightarrow 0$ and $H \rightarrow H_c$, where the $q_2$ order collapses while the residual Sommerfeld coefficient $\gamma_0$ remains finite, indicating a large density of low-energy excitations coexisting with long-range magnetic order. 
Correspondingly, the low-energy gap $\Delta_L$ closes at $H_c$, whereas the higher-energy gap $\Delta_H$ remains finite within Phase~II near $H_c$ [Fig.~\ref{fig:phasecontour}(b,c)].

This coexistence of a finite $\gamma_0$ and partially gapped excitations provides strong evidence for a mode-selective quantum critical point driven by the selective softening of the $q_2$ branch. 
Polarized neutron diffraction corroborates this picture, showing suppression of the $q_2$ component near $1.7$--$2$~T while the $q_1$ order remains robust up to 6~T for $H||[110]$~\cite{Marshall2022}. 
The resulting magnetic state retains long-range order yet hosts enhanced low-energy spin fluctuations as $T \rightarrow 0$. Within this regime, we extract effective critical exponents significantly smaller than those in conventional phase transitions, consistent with criticality confined to a restricted portion of phase space. 
The finite $\gamma_0$ further indicates unconventional low-energy excitations beyond a standard magnon description, suggesting a possible change in their nature. 
Further microscopic probes will be required to elucidate the underlying spin dynamics.

BaNd$_2$ZnS$_5$ thus provides a platform for selective quantum criticality in frustrated rare-earth magnets, where low-energy excitations reorganize into coexisting gapped and gapless modes. 
This behavior establishes a route to field-tunable quantum criticality in strongly spin–orbit–entangled systems, enabled by symmetry-selective coupling in $4f$ materials.

\section*{Acknowledgements} \label{sec:acknowledgements}
This material is based upon work supported by the US Department of Energy, Office of Science, National Quantum Information Science Research Centers, Quantum Science Center. M. L., S. L., S. Z, and V. Z. were funded by QSC to perform data analysis and manuscript writing. C. A. M.,  B. M., and M. L. acknowledge the support of the Laboratory Directed Research and Development program of Los Alamos National Laboratory under Project No. 20240225ER. A portion of this work was performed at the National High Magnetic Field Laboratory, which is supported by National Science Foundation Cooperative Agreement No. DMR-2128556, the State of Florida, and the U.S. Department of Energy.
The work at the University of Arizona was supported by the U.S. Department of Energy (DOE), Office of Science, Basic Energy Sciences (BES) under Award DE-SC0025301.

\bibliographystyle{naturemag} 
\bibliography{Reference}

\end{document}